\documentclass[fleqn,10pt]{wlscirep}

\usepackage[utf8]{inputenc}
\usepackage[T1]{fontenc}
\usepackage{enumitem}
\usepackage{amsmath,amssymb,amsfonts,mathtools,upgreek}
\usepackage{bm,bbm}
\usepackage[squaren]{SIunits}
\usepackage{graphicx,float,caption}
\usepackage{hyphenat}
\usepackage{multirow}

\usepackage[font=small,
labelfont=bf,
justification=justified,
format=plain]{caption}

\title{Synchronous, Crosstalk-free Correlative AFM and Confocal Microscopies/Spectroscopies}

\author[1]{Thales F. D. Fernandes}
\author[1]{Oscar Saavedra V.}
\author[1]{Emmanuel Margeat}
\author[1*]{Pierre-Emmanuel Milhiet}
\author[1**]{Luca Costa}

\affil[1]{Centre de Biochimie Structurale (CBS), CNRS, INSERM, Univ Montpellier, 34090, Montpellier, France}
\affil[*]{pem@cbs.cnrs.fr}
\affil[**]{costa@cbs.cnrs.fr}

\begin{abstract}
    Microscopies have become pillars of our characterization tools to observe biological systems and assemblies.
	Correlative and synchronous use of different microscopies relies on the fundamental assumption of non-interference during images acquisitions.
	In this work, by exploring the correlative use of Atomic Force Microscopy and confocal-Fluorescence-Lifetime Imaging Microscopy (AFM-FLIM), we quantify cross-talk effects occurring during synchronous acquisition.
	We characterize and minimize optomechanical forces on different AFM cantilevers interfering with normal AFM operation as well as spurious luminescence from the tip and cantilever affecting time-resolved fluorescence detection.
	By defining non-interfering experimental imaging parameters, we show accurate real-time acquisition and two-dimensional mapping of interaction force, fluorescence lifetime and intensity characterizing morphology (AFM) and local viscosity (FLIM) of gel and fluid phases separation of supported lipid model membranes.
	Finally, as proof of principle by means of synchronous force and fluorescence spectroscopies, we precisely tune the lifetime of a fluorescent nanodiamond positioned on the AFM tip by controlling its distance from a metallic surface.
	This opens up a novel pathway of quench sensing to image soft biological samples such as membranes since it does not require tip-sample mechanical contact in contrast with conventional AFM in liquid.
\end{abstract}

\begin{document}

\flushbottom
\maketitle
\thispagestyle{empty}

\section*{Introduction} \label{sec:intro}
Real space observation of biomolecules at high resolution is crucial to address fundamental biological questions.
In this frame, several microscopy techniques have been developed during the last decades and are currently well-established and available allowing characterizations at different lengths and timescales.
Cryogenic Electron Microscopy (cryo-EM) offers nowadays the highest spatial resolution at the molecular and sub-molecular length scales but cannot image living cells in physiological conditions~\cite{Cheng2018}.
For that, a large variety of optical-based microscopies are available, historically less resolved than cryo-EM because of the diffraction limit.
However, some recently developed super-resolution techniques such as Stimulated Emission Depletion Microscopy (STED)~\cite{Willig2006}, Stochastic Optical Reconstruction Microscopy (STORM)~\cite{Rust2006}, Photoactivated Localization Microscopy (PALM)~\cite{Manley2008} and MINFLUX~\cite{balzarotti2017nanometer} have bypassed the diffraction limit and improved the lateral resolution down to a few nanometers.
While they do have some drawbacks both in terms of photo-toxicity or acquisition speed, their popularity nowadays reflects their coming of age.
When it is to observe bio-interfaces at high resolution in physiological conditions, in a label-free configuration, Atomic Force Microscopy (AFM)~\cite{Dufrene2017} and its fast video-rate version High-Speed AFM (HS-AFM)~\cite{Ando2012,Nasrallah2019} are outstanding techniques to obtain morphological and mechanical information at nano-scale.
While all microscopies provide complementary physical and chemical information at different spatial and temporal resolutions, the challenge is now to get correlative observations of biological processes with two or more techniques that require important technological efforts.

As examples, cryo-EM and 3D super-resolution fluorescence have been coupled to aid protein localization in EM images~\cite{Kopek2012}, as well as STED~\cite{Harke2012}, STORM~\cite{Odermatt2015,Dahmane2019a} and Super-Resolved Structured Illumination Microscopy (SR-SIM)~\cite{Gomez-Varela2020} have been successfully coupled with AFM. 
Combining conventional fluorescence and AFM has become increasingly popular since both techniques can work in biologically relevant conditions and they acquire complementary information: specificity mapping of proteins given by specific fluorescent probes nicely complements the high-resolution AFM images.
Confocal Laser Scanning Microscopy (CLSM) and AFM have been coupled at first in 1995~\cite{Hillner1995} to image Langmuir-Blodgett films of 10,12-pentacosadiynoic acid.
In the following decades, a large number of setups following similar operational schemes based on fixed optical axes, fixed tip position, and a scanning sample were developed~\cite{Gradinaru2004,Kassies2005,Bek2008,Tisler2013,Tetienne2014b,he2016simultaneous}.
In most setups, AFM and fluorescence images are acquired in a sequential manner, and correlation between images is generated afterwards~\cite{Meller2006,Laskowski2017}.
However, synchronicity is of high importance as it permits to follow dynamical biological processes.

In addition, it allows the acquisition of optical images while a force is applied by the AFM tip in pump and probe experiments, for instance, monitoring the response of a living cell to an AFM tip indentation or during single-molecule manipulation~\cite{Beicker2018,Efremov2019}.
This simultaneous modality is often not allowed due to interference and cross-talk during simultaneous image acquisition resulting in worse performance and data quality, both effects often ignored in the existing literature and only very rarely discussed~\cite{Miranda2015,he2016simultaneous}.
When coupling fluorescence (in both wide-field~\cite{Cazaux2016} and confocal~\cite{he2016simultaneous} excitation schemes) with AFM, it is, therefore, of critical importance to eliminate any cross-talking between the two techniques by understanding, quantifying, and reducing the artifacts occurring during synchronous operation.
We can divide artifacts into two categories:
\begin{enumerate}
	\item AFM acquisition being affected by the excitation laser used to excite fluorescent probes.
	The interaction of the laser confocal spot and the AFM tip/cantilever could give rise to optomechanical/photothermal forces that can deflect the cantilever in a non-linear way which depends upon the distance from the sample.
	Since in time-resolved fluorescence experiments the excitation laser is continuous or pulsed at operating frequencies in the MHz regime, typically much higher than the most of cantilever resonance frequencies, the measured force (the deflection) is due to the static component of the excitation.
	\item Fluorescence acquisition being affected by the presence of the AFM tip/cantilever.
	AFM tip luminescence gives rise to a spurious signal in the photons detection that might not be fully eliminated if there is an overlap between tip/cantilever fluorescence and the sample emission spectra.
\end{enumerate}

In the case of wide-field epifluorescence, the optomechanical interactions between light and the cantilever body can result in large cantilever deflections. Cazaux \emph{et al.} have used these deflections as a mean to mechanically synchronize AFM and optical measurements~\cite{Cazaux2016}.
A wide-field excitation scheme such as Total Internal Reflection excitation (TIRF) eliminates most artifacts since it limits the interaction of light only to a few hundred nanometers above the glass coverslip, where solely the tip's end is present, the cantilever being typically few microns axially farther. A large number of simultaneous experiments were reported in the last two decades using this modality~\cite{Sarkar2004,Gumpp2009,Ortega-Esteban2015,Uchihashi2016,Umakoshi2019}.
However, this type of illumination is not suitable for thick samples such as cells, since the tip will be positioned at the apical cellular membrane axial position, out of the evanescent excitation field.

Confocal microscopy is perfectly suited for such cellular imaging, but the interaction of the AFM tip and the confocal spot can, unfortunately, extend to a greater distance.
In this work, we focus on the coupling between a confocal spot and an AFM tip/cantilever with the aim to investigate and quantify photothermal induced deflection resulting from the presence of metal coating at the cantilever backside~\cite{Ramos2006}, radiation pressure exerted on the tip~\cite{Favero2009} due to the scattering of the confocal spot and finally tip/cantilever luminescence~\cite{Kassies2005}.
We show that all effects highly depend on the choice of the AFM tip and cantilever.
We conclude that tip geometry, material, and metallic coating at the cantilever backside, as well as cantilever stiffness, are the key parameters to be taken into account.
We also present how these common pitfalls can be avoided, or highly minimized, so there is no spurious cross-talk in the acquired images.
As proof of principle we imaged synchronously the topography, lifetime, and fluorescence intensity of gel and fluid phases in Supported Lipid Bilayer (SLB), labeled with a BODIPY fluorescent probe.
Finally, for the first time, to the best of our knowledge, we implemented a simultaneous AFM Force Spectroscopy (AFM-FS) and confocal fluorescence spectroscopy operational scheme.
In this way, we probed the force and lifetime variation due to Metal Induced Energy Transfer (MIET) between the Nitrogen-Vacancy (NV) centers in a nanodiamond attached to the apex of an AFM tip and a substrate of thin gold, therefore, enabling a novel pathway of quench sensing and opening up novel non-invasive imaging modes in liquid that can be used to image soft and fragile biological specimens.

\section*{Results} \label{sec:results}
\begin{figure}[htb]
	\includegraphics[width=10cm]{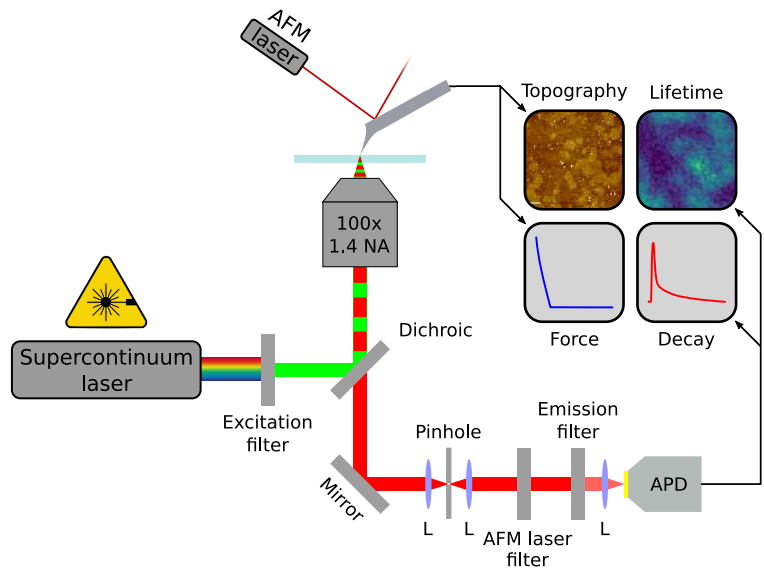}
	\centering
	\caption{
		AFM and inverted confocal microscope setup, where the AFM and the optics are coupled by synchronizing the photons acquisition with a TTL signal from the AFM electronics controller.
		Convergent lenses are indicated as L.
	}
	\label{fig:setup}
\end{figure}

The synchronous imaging acquisition setup based on confocal excitation is described by Fig.~\ref{fig:setup}.
The excitation source is a supercontinuum pulsed laser where an excitation band-pass filter enables the convenient switching between different excitation wavelengths for multi-color fluorescence acquisition.
The light is focused by a high Numerical Aperture objective (NA = $1.4$).
The emitted fluorescence photons are collected by the objective, focused on a pinole, recollimated, filtered by an emission filter and finally focused on an Avalanche Photodiode (APD) connected to a Time-Correlated Single Photon Counting (TCSPC) card.

The AFM is mounted on top of an inverted microscope where the sample and the tip can be displaced independently with two XYZ~piezo scanners.
The synchronization of the AFM and the TCSPC card is made by routing the events (photons) with a Transistor-Transistor Logic (TTL) signal generated by the AFM electronics controller, therefore, enabling to assign a $xy/z$-position for each photon.
Topography, intensity, and lifetime signals from the setup are used to construct simultaneously an AFM and two optical images, respectively.
If operated in force spectroscopy mode, (AFM-FS), force, intensity, and lifetime curves versus tip-sample distance are simultaneously acquired.
The lifetime was either evaluated for each image pixel by fitting a decay curve with a single exponential or, in AFM-FS mode, by amplitude weighting the decay curve (see Materials and Methods section).
It is worth to notice that is crucial to use a short-pass filter (AFM laser filter) before the APD to filter out the light used in the AFM cantilever monitoring setup (AFM laser)~\cite{Araghi2017}, usually called optical beam deflection.

\begin{figure}[htb]
	\includegraphics[width=14cm]{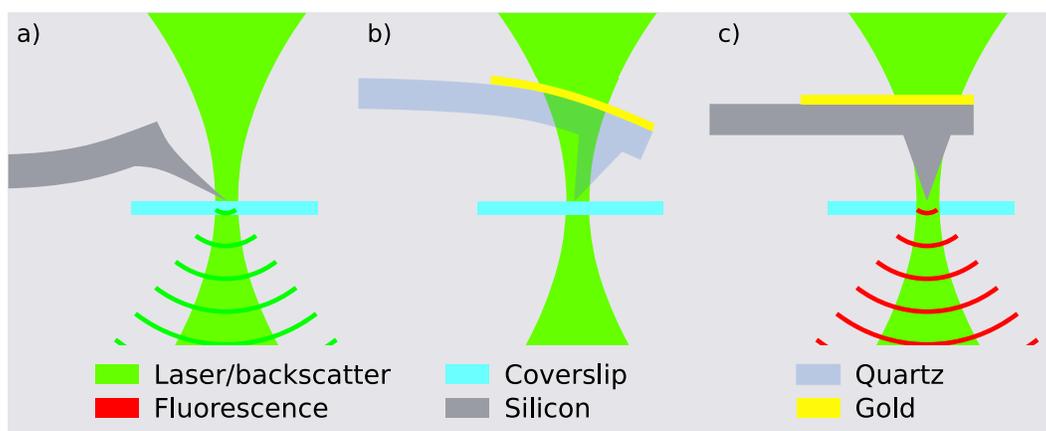}
	\centering
	\caption{
		Possible AFM tip and confocal spot interactions: radiation pressure (\textbf{a}), photothermal induced deflection (\textbf{b}), and cantilever/tip luminescence (\textbf{c}).
	}
	\label{fig:forces}
\end{figure}

As previously mentioned, the interaction of the focused excitation light and the AFM tip can give rise to two different phenomena: optomechanical forces and/or tip/cantilever luminescence, where the former branches into radiation pressure and photothermal effect.
Radiation pressure (Fig.~\ref{fig:forces}a) will always be present in all cases since the scattering of the incoming light will generate forces due to change in momentum of light.
Momentum conservation dictates that the cantilever will be deflected upwardly by this force.
Photothermal deflection (Fig.~\ref{fig:forces}b) is present if the cantilever is composed of two different materials (similarly to a bimetallic strip), which is the case of metal-coated cantilevers, where a thin layer of metal is deposited onto the cantilever backside to improve reflectivity.
For visible light, gold will absorb more than either silicon or silicon nitride (higher extinction coefficient) and, due to its higher thermal expansion coefficient, it will expand more when heated by the laser.
This differential expansion will cause a downwards bending of the cantilever.
Spurious luminescence (Fig.~\ref{fig:forces}c) can arise from two sources: 1. the material of the tip/cantilever itself can present luminescence properties, as in the case of amorphous silicon nitride, particularly important at wavelengths higher than $800$~nm~\cite{Kassies2005}. 2. The coating on the cantilever backside can be luminescent as well, as for example gold that emits in the red region of the spectrum~\cite{YungHui2013}.
It is worth to notice that backscattered light will occur in all possible configurations since the tip will always scatter the incoming light of the confocal spot in all directions and, therefore, for the same reason radiation pressure will always be present.
However, most of the backscattered light can be easily eliminated by a notch filter when it is not already sufficiently reflected by the emission filter.

\begin{figure}[htb]
	\includegraphics[width=16cm]{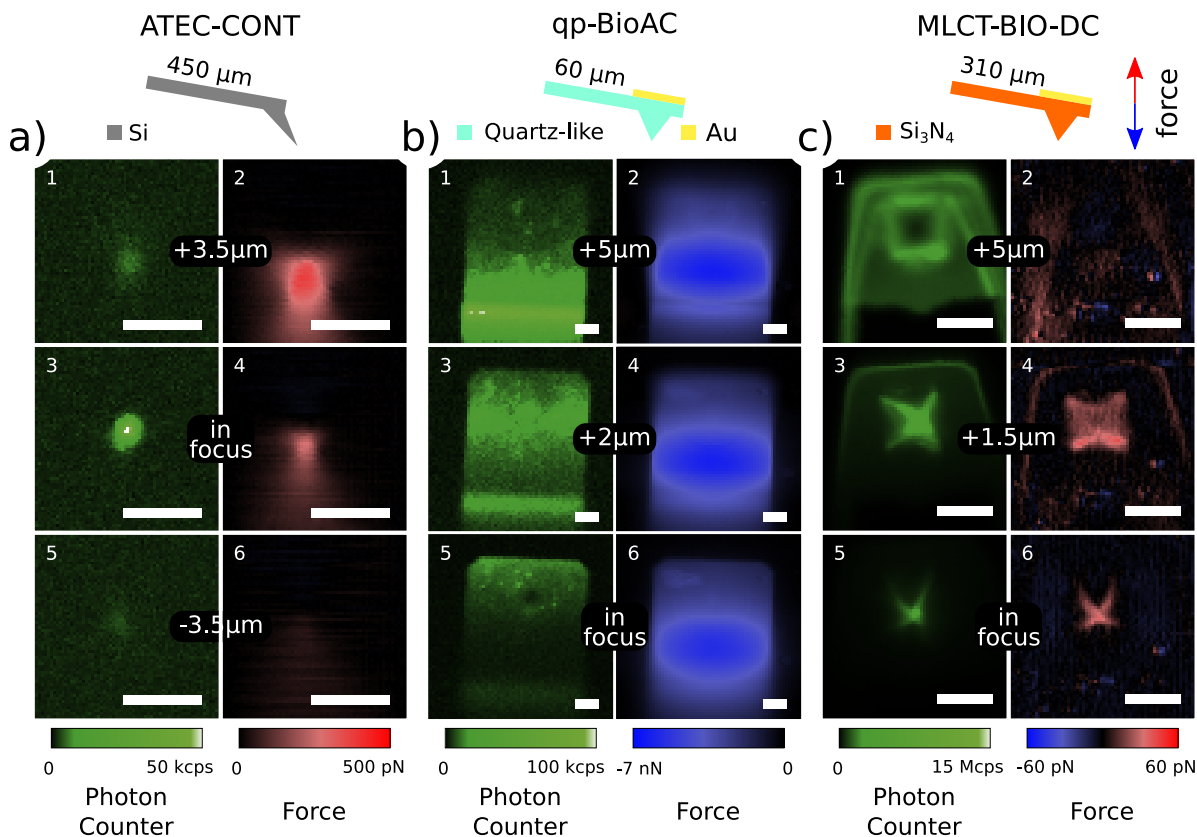}
	\centering
	\caption{
		Simultaneous photon counts (panels 1, 3, and 5) and force (panels 2, 4, and 6) scans for three different AFM cantilevers: (\textbf{a}) ATEC-CONT, (\textbf{b}) qp-BioAC, and (\textbf{c}) MLCT-BIO-DC.
		The images were acquired using a reverse-tip image operational scheme where the confocal spot is a probe and the AFM tip and cantilever are samples to be measured.
		Data were collected with a $532$/$10$~nm excitation filter, a $690/50$~nm emission filter, and a laser power of $300~\upmu$W.
		Subfigures 1-2, 3-4, and 5-6 were acquired at different distances from the confocal spot focal plane.
		The scale bar is $5~\upmu$m for all images.
		In the schematic of the cantilevers, the nominal cantilever length and the red and blue arrows indicate positive and negative force directions, respectively, corresponding to the color scale in panels 2, 5, and 6.
		Intensity (photon counter) images (panels 1, 3, and 5) are given in counts per second (cps) and were recorded directly from the APD.
		The cantilever used from the qp-BioAC is the CB2 while the one for MLCT-BIO-DC is the cantilever C. The Point Spread Function (PSF) of the confocal spot is $\approx 400$~nm.
	}
	\label{fig:z-scan}
\end{figure}

We characterized and quantified these effects by operating the AFM in the reverse-tip imaging configuration~\cite{Rondin2012}.
The confocal spot is focused $4\upmu$m from a glass coverslip and held in place throughout the measurement.
Using the \textit{hover mode} of the AFM (see Material and Methods), the tip scans the glass coverslip in the trace scan and then, in the retrace scan, the tip follows the retrace topography vertically displaced in $z$ by a given distance specified by the user.
In this way, it is possible to make a constant height image even if the sample is inclined.
In this mode, the deflection of the cantilever is monitored in the retrace scan without an active feedback.
Deflection (force) and intensity (photon counter) images are acquired during scan, as shown in Fig.~\ref{fig:z-scan}.
The resulting images will be of the interaction of the confocal spot with different regions of the tip for different focal planes.
In this operational scheme, the confocal spot is steady and operates as a ``probe'' while the AFM tip and cantilever are the samples to be measured, hence the name, reverse-tip imaging.
We used three AFM cantilevers with different specifications: a protruding silicon tip (Fig.~\ref{fig:z-scan}a), a quartz-like tip with gold coating on the cantilever backside (Fig.~\ref{fig:z-scan}b) and silicon nitride (Si$_3$N$_4$) with a gold-coated cantilever (Fig.~\ref{fig:z-scan}c), commercially available as ATEC-CONT (Nanosensors), qp-BioAC (Nanosensors), and MLCT-BIO-DC (Bruker), respectively.
The images were acquired both with the tip in focus and out-of-focus positions to observe the different interactions of the confocal spot ($300~\upmu$W power, $532/10$~nm wavelength) with the AFM tip and cantilever body.

Given the large emission spectra of silicon~\cite{silicon}, silicon nitride~\cite{Kassies2005}, and the metallic coating on the cantilever backside, in our case gold~\cite{YungHui2013}, the choice of the emission filter is not crucial if compared to fluorescence imaging of conventional organic fluorophores characterized by narrow emission spectra.
In Fig.~\ref{fig:z-scan} we used a $690/50$~nm emission filter.
Panels 1, 3, and 5 show the photon intensity signal whereas panels 2, 4, and 6 display the vertical force acting on the AFM cantilever.
Fig.~\ref{fig:z-scan}a shows that the ATEC-CONT has a bright central spot in the photon intensity channel that is due to tip luminescence and presents a positive force in the range of a few hundred of piconewtons.
By inserting a notch filter in the emission path, no significant change in the photon intensity channel is observed, which shows that backscattering contribution is minimal and it is completely filtered out by the band-pass emission filters.
Such a positive force is a direct consequence of the radiation pressure acting on the cantilever due to the backscattered light.
With the qp-BioAC (Fig.~\ref{fig:z-scan}b) the situation is consistently different: the force is negative and it has a maximum modulus of $\approx7~$nN.
The confocal spot interacts strongly with the gold reflective coating on the cantilever backside, as shown in the photon intensity images, Fig.~\ref{fig:z-scan}b panels 1, 3, and 5.
As a consequence, the interaction induces a differential heating that bends the cantilever downwards due to photothermal effects.
The intensity signal presents no direct correlation with the force observed and is highly dependent on the cantilever bending angle.
In contrast with the previous case, the tip position is characterized by a darker spot in the photon intensity image (Fig.~\ref{fig:z-scan}b panel 5) that is due to the fact that both excitation light and gold emission have to pass through the tip, therefore through a higher quantity of material.
As a consequence, we collect less fluorescence emitted by the reflective gold coating material when the tip and confocal spot are aligned.
The MLCT-BIO-DC (Fig.~\ref{fig:z-scan}c) shows a mix of radiation pressure and a mild photothermal deflection contributions in the force signal.
In this case, it is also possible to observe the interaction with the tip's gold-coating but the thermal effects are not as drastic as before, leading to a positive total force acting on the cantilever when the tip and confocal spot are aligned (see Materials and Methods).
The small deflection of the MLCT-BIO-DC, despite the presence of a gold coating at the cantilever's end, could be attributed to its manufacturing process, optimized to minimize probe bending due to elevated temperatures.

\begin{figure}[htb]
	\includegraphics[width=8.1cm]{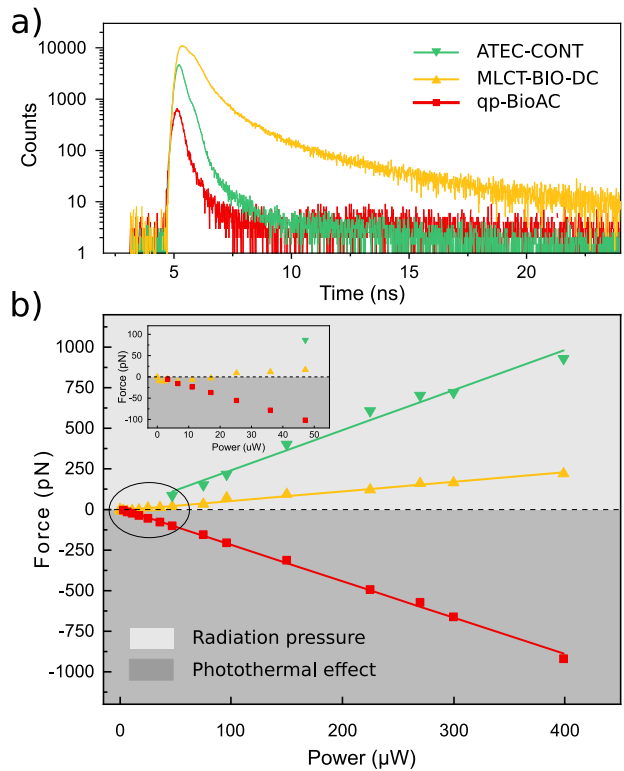}
	\centering
	\caption{
		(\textbf{a}) Luminescence decay curves for different AFM cantilevers and (\textbf{b}) variation of the measured force as a function of the laser power.
		The cantilevers are qp-BioAC (red), MLCT-BIO-DC (yellow), and ATEC-CONT (green).
		The inset in (\textbf{b}) shows a zoom of the region of low power, highlighted by the ellipse.
		Positive force is characterized as radiation pressure while negative force is photothermal induced deflection.
		The slopes of the lines in (\textbf{a}) are: ATEC-CONT ($2.47\pm0.13$~pN/$\upmu$W), MLCT-BIO-DC ($0.59\pm0.02$~pN/$\upmu$W), and qp-BioAC ($-2.24\pm0.03$~pN/$\upmu$W).
		All curves were acquired with a $532$/$10$~nm excitation filter, with a $690/50$~nm emission filter, and $300~\upmu$W power in~(\textbf{a}).
		The positive force region is colored in green to indicate the radiation pressure regime whereas the negative force region is colored in purple to indicate the photothermal regime. 
	}
	\label{fig:PL-force}
\end{figure}

Given the data presented in Fig.~\ref{fig:z-scan}, the arising question is what cantilever is worth using to reduce cross-talk between AFM and time-resolved fluorescence acquisitions in correlative experiments.
Fig.~\ref{fig:PL-force}a presents time-resolved fluorescence decay curves acquired for the three AFM cantilevers of Fig.~\ref{fig:z-scan} with a high laser power of $300~\upmu$W to maximize signal to noise ratio for the less luminescent ones.
MLCT-BIO-DC (yellow curve) has the highest intensity signal by two orders of magnitude and presents a long lifetime in comparison with the qp-BioAC (red curve) and the ATEC-CONT (green curve).
While ATEC-CONT and qp-BioAC will affect the measured lifetime of fluorescent probes characterized by a short ($\approx1$~ns) lifetime, the MLCT-BIO-DC will affect a larger variety of fluorophores presenting even longer lifetimes.
In addition, we monitored the variation of the spurious force as a function of the incident laser power (Fig.~\ref{fig:PL-force}b).
The force scales linearly with the power and the different cantilevers branch in the two categories depending on the nature of the optomechanical interaction: radiation pressure and photothermal induced deflection.

It is worth to notice that the photothermal deflection computed in this case, for a laser power of $300~\upmu$W, is inferior in modulus to the $7$~nN shown in Fig.~\ref{fig:z-scan}b because in the latter case the maximum force is measured with the confocal spot focused on gold-coated cantilever backside and not on the AFM tip as in Fig.~\ref{fig:PL-force}b, as previously mentioned.
Table~\ref{tab:AFMcant} reports the slopes from Fig.~\ref{fig:PL-force}b, force versus laser power, which indicates the significance of the force for a given probe.
\begin{table}[h] 
    \caption{Slopes of the fits in Fig.~\ref{fig:PL-force}}
    \centering
     \begin{tabular}{c c c}
        \hline
        \multicolumn{1}{c}{\multirow{2}{*}{Cantilever}} & \multirow{2}{*}{Brand} & Optomechanical \\
        \multicolumn{1}{c}{} & & constant [~pN/$\upmu$W ] \\
        \hline 
        ATEC-CONT & Nanosensors & \phantom{$-$}$2.47\pm0.13$~pN/$\upmu$W \\
        qp-BioAC & Nanosensors & $-2.24\pm0.03$~pN/$\upmu$W \\
        MLCT-BIO-DC & Bruker & \phantom{$-$}$0.59\pm0.02$~pN/$\upmu$W \\
        \hline
    \end{tabular}
    \label{tab:AFMcant}
\end{table}

Concluding, low laser power should be used, and the relative position between the tip and the confocal spot should be held constant, leading to a change of the cantilever equilibrium position while avoiding a change of relative deflections.
In addition, our data suggest the qp-BioAC cantilevers should be employed to accurately evaluate the lifetime of the fluorescent probes within the sample (Fig.~\ref{fig:PL-force}a).
Instead, MLCT-BIO-DC cantilevers, presenting lower spurious force, are more suited for experiments where low AFM interaction forces have to be measured and in presence of a large number of fluorophores in the excitation volume, ensuring high signal to noise ratio to properly evaluate their lifetime.

\begin{figure}[htb]
	\includegraphics[width=16cm]{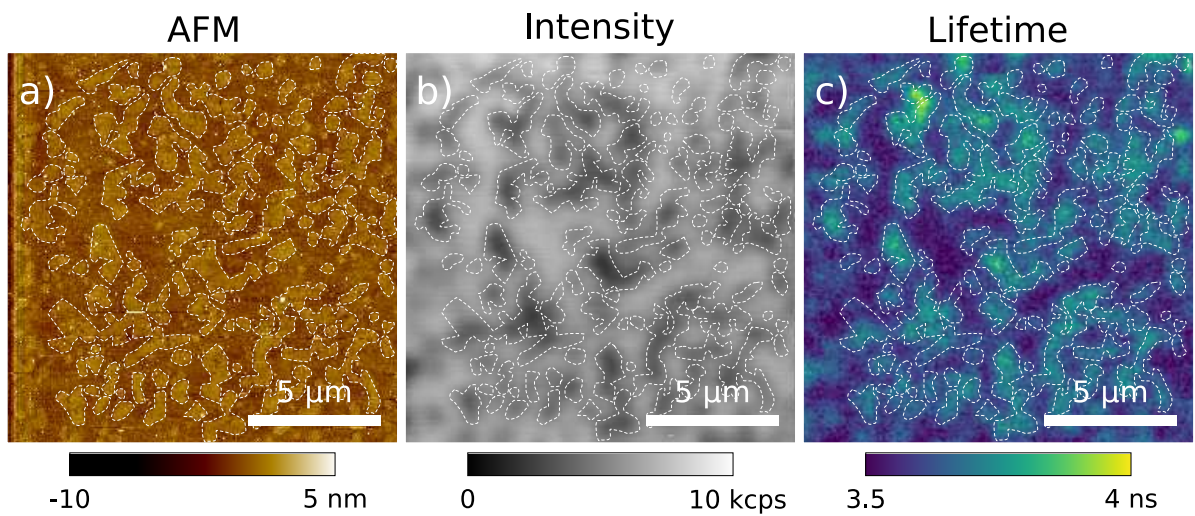}
	\centering
	\caption{
		AFM morphology (\textbf{a}), fluorescence intensity (\textbf{b}), and lifetime (\textbf{c}) of DOPC:DPPC ($1:1$ molar ratio) labeled with $0.1\%$ BODIPY model membranes.
		The three images were synchronously acquired.
		The scale bar is $5~\upmu$m, the power used was $120$~nW with a $488/10$~nm excitation filter and a $525/39$~nm emission filter.
		The tip used was a qp-BioAC.
		The dashed white line serves as visual aid for the separation of the two domains.
	}
	\label{fig:SLB}
\end{figure}

The advantage of the correlative AFM-FLIM imaging mode relies on the complementarity of the two techniques.
To prove that, we imaged fluid and gel phases enriched domains in biological model membranes, correlating their morphology (AFM) with their local viscosity (FLIM).
We used a lipid mixture of 1,2‐dipalmitoyl‐sn‐glycero‐3‐phosphocholine (DPPC), 1,2-dioleoyl-sn-glycero-3- phosphocholine (DOPC) in $1:1$ molar ratio supplemented with $0.1\%$ in volume of 2-(4,4-difluoro-5,7-dimethyl-4-bora-3a,4a-diaza-s-indacene-3-dodecanoyl)-1-hexadecanoyl-sn-glycero-3-phosphocholine (BODIPY).
For the simultaneous acquisition of AFM, confocal microscopy, and FLIM (shown in Fig.~\ref{fig:SLB}), we used the qp-BioAC cantilever and a laser power of $120$~nW, ensuring, as previously explained, the highest accuracy for BODIPY lifetime evaluation while minimizing the spurious force during AFM acquisition.

Both topography (Fig.~\ref{fig:SLB}a) and fluorescence intensity (Fig.~\ref{fig:SLB}b) show that the DPPC enriched regions (higher thickness) have lower BODIPY concentration.
Wu~\emph{et al.} showed in Giant Unilamellar Vesicles (GUVs) that high viscosity lipids present higher lifetime, hence lifetime can be used to probe directly the lipids viscosity~\cite{Wu2013}.
Fig.~\ref{fig:SLB}c shows that DPPC enriched domains have higher lifetime when compared with DOPC domains, this is expected since DPPC is a gel-phase lipid, hence more viscous than DOPC, a fluid-phase lipid.
The higher lifetime, above $3$~ns, when compared to Wu~\emph{et al.} $\approx1.8$~ns~\cite{Wu2013}, can be attributed to the fact that supported bilayers are red on top of a substrate (glass), inducing different physical behaviors in terms of lipid diffusion and potentially local viscosity when compared to free-standing bilayers in the GUVs case. Indeed, even if the presence of a layer of buffer (2-3 nm thickness) between the lipid polar heads and the substrate generally prevents an important hindrance in lipid diffusion \cite{SEANTIER2008326}, it is known that the latter can be reduced as compared to giant unilamellar vesicles \cite{Hof}.
The tip luminescence, dominated by low lifetime ($<1$~ns) is not present in the simultaneous AFM-FLIM imaging (see sections~3 and 5 of the supplementary information).

Minimizing tip/cantilever luminescence and optomechanical forces is relevant also in synchronous AFM-FS and fluorescence spectroscopy experiments.
In this novel operational scheme, the AFM tip has fluorescent properties that can be correlated with the force measured by AFM while being approached or retracted from the sample.
The force readout can be used to define the mechanical contact position between the sample and the fluorescent probe and, subsequently, the latter can be accurately placed with nanometric precision at a given distance from the sample and its lifetime can be probed.

\begin{figure}[htb]
	\includegraphics[width=16cm]{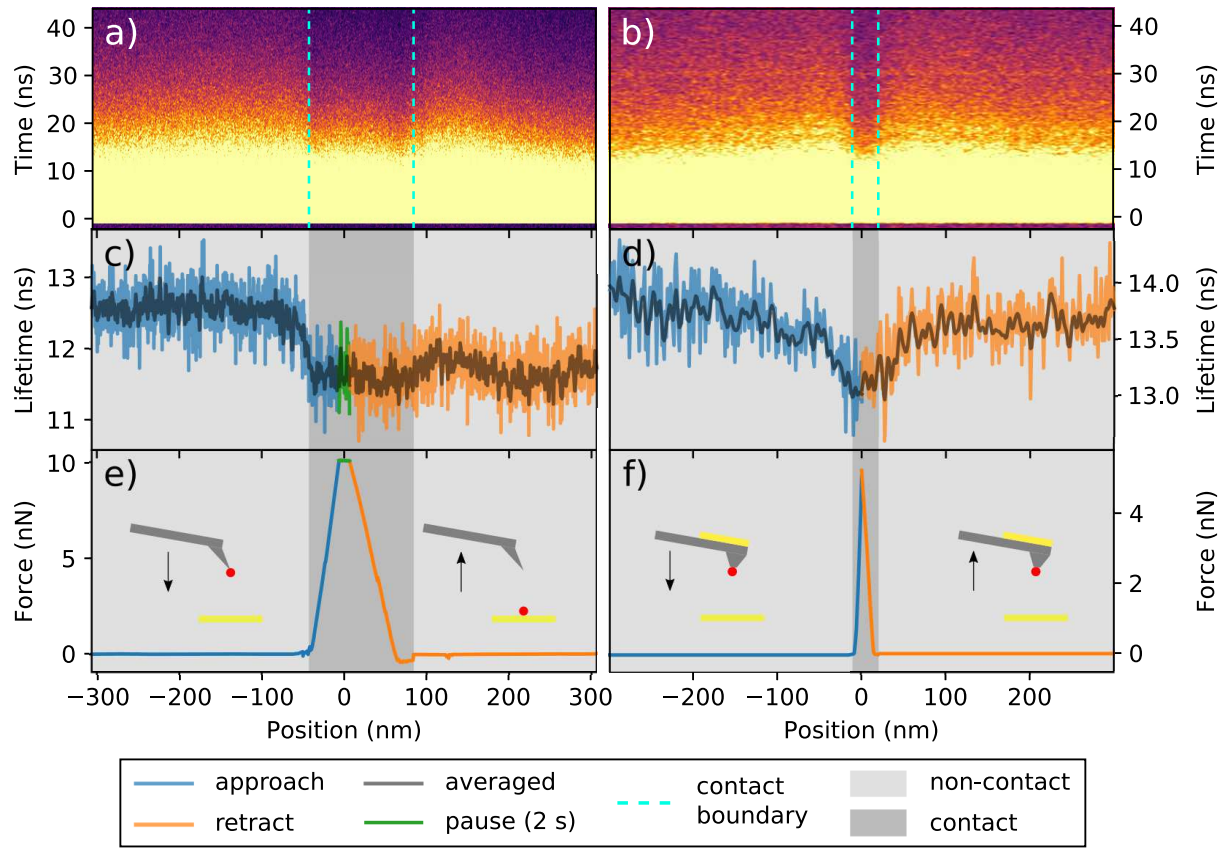}
	\centering
	\caption{
		Simultaneous force curve and nanodiamond lifetime measurements approaching and retracting a gold substrate.
		Photons are synchronized with TTL from the AFM that enables the acquisition of decay curves for every $z$-position of the tip.
		(\textbf{a} and \textbf{b}) show the tip $z$-position versus time (decay curves), where the color scale represents the number of photons collected.
		(\textbf{c} and \textbf{d}) show the lifetime of the nanodiamond in function of the tip position.
		(\textbf{e} and \textbf{f}) show the standard force curves unfolded so approach and retract curves are separated in the negative and positive axis, respectively.
		The blue curves in (\textbf{c}-\textbf{f}) consists of the approach whereas orange region the retract.
		The green curves in (\textbf{c} and \textbf{e}) constitute a region where the tip was paused for $2$~seconds.
		The light gray region displays the non-contact (zero force) interactions whereas the darker gray consists of the contact regime.
		A $532/10$~nm excitation filter was used with a $709/167$~nm emission filter with power of $90~\upmu$W.
		The dark gray curves in (\textbf{c} and \textbf{d}) show the data with spatial ($9$~neighbors) and temporal ($4$~neighbors) binning (filter).
		The same binning is also applied to (\textbf{a} and \textbf{b}) to aid visualization.
		An ATEC-CONT was used in (\textbf{a}, \textbf{c}, and \textbf{f}) whereas in (\textbf{b}, \textbf{d}, and \textbf{f}) a modified qp-BioAC red(CB1) with a plateau was used.
		Tip schematics in \textbf{e} and \textbf{f} show an AFM tip with a nanodiamond (in red) and the gold substrate (yellow).
		In the retract curve of (\textbf{e}) the nanodiamond is left on the surface whereas in (\textbf{f}) it is kept at the tip's end.
	}
	\label{fig:force-lifetime}
\end{figure}

To demonstrate that, we show the decrease of lifetime of NV~centers in a diamond nanoparticle due to MIET~\cite{Drexhage1970} while the nanodiamond is approached to a gold substrate.
Fig.~\ref{fig:force-lifetime} shows an example of nanopositioning and nanomanipulation of a nanodiamond attached to the apex of an AFM tip.
Figs.~\ref{fig:force-lifetime}a and b show a plot of the decays for every pixel ($z$-position, axis), where the color scale represents the number of photons collected, acquired simultaneously with a force curve Fig.~\ref{fig:force-lifetime}e and~f.
The experiments were performed in Dulbecco's Phosphate-Buffered Saline (DPBS) solution.
Firstly, we approached an ATEC-CONT tip holding a nanodiamond at its apex to a gold substrate, pressed with a high force ($10$~nN) and held in place for $2$~seconds (Figs.~\ref{fig:force-lifetime}a, c, and e).
This allowed the nanodiamond to dislodge from the tip and to stay permanently on the gold.
The tip was then retracted without the nanodiamond.
This can be clearly seen from Fig.~\ref{fig:force-lifetime}c, where the lifetime decreases approaching the gold and never recovers back during tip retraction from gold.
The lifetime of the nanodiamond changes from $12.60\pm0.28$~ns to $11.69\pm0.30$~ns, which corresponds to a quenching efficiency of $\approx7.2$~\%.
In a second experiment, showed in Figs.~\ref{fig:force-lifetime}b, d, and f, we used a modified qp-BioAC (CB1) with a plateau at its end holding a stable nanodiamond at the tip's apex (see Materials and Methods section).
In this configuration, the fluorescent nanodiamond is approached, pressed with lower force ($5$~nN), and retracted from the gold substrate without being dislodged from the AFM tip.
The lower force is a requirement to avoid the loss of the nanodiamond upon contact.
The lifetime and force curves are symmetric, Figs.~\ref{fig:force-lifetime}d and f, showing that the interaction is reversible.
The lifetime of the nanodiamond changes from $13.86\pm0.27$~ns to $13.13\pm0.23$~ns upon contact with the gold, which corresponds to a quenching efficiency of $\approx5.5$~\%.
The quenching in lifetime when approaching gold, or any metal, is expected due to MIET~\cite{Drexhage1970}.
The radiation rate will change due to the coupling of the nanodiamond electromagnetic near-field with the plasmons of the gold.
In the vertical position ($x$-axis) of Figs.~\ref{fig:force-lifetime}e and f the approach and retract curves have been ``unfolded'', where approach/retract have negative/positive position values.
A folded version of the force curves is presented in section~2 of the supplementary information together with the lifetime curve for another nanodiamond.

\section*{Discussion} \label{dic}
The mutual interaction of the confocal spot and the AFM tip/cantilever can give rise to a cross-talk in the AFM and/or fluorescence images.
Two main forces have been explored: radiation pressure and photothermal-induced deflections.
Miranda~\emph{et al.} have calculated the deformation due to laser heating on gold-coated cantilevers and found displacements of $9$~nm and forces around $3$~nN with $\approx2$~times less power than used in Fig.~\ref{fig:z-scan}~\cite{Miranda2015}.
Radiation pressure forces can be estimated to be in the range of few tens/hundreds of piconewtons, in agreement with our results (see section~1 of the supplementary information).
While in standard confocal imaging a laser power inferior to $100~\upmu$W is usually employed, in correlative STED-AFM~\cite{Odermatt2015} a much higher power in the mW range is required, suggesting that simultaneous STED and AFM will produce optomechanical forces from several hundreds of pN to the nN range (Fig.~\ref{fig:PL-force}b), causing large tip deflections when using soft AFM cantilevers ($<0.1$~N/m stiffnesses).
The inset of Fig.~\ref{fig:PL-force}b, however, suggests that even at reasonably low laser power ($10$-$50~\upmu$W) optomechanical forces are present and can be quantified in few tens of pN.
The same applies to the overall attractive force in the case of qp-BioAC.
Those forces are not negligible and should be taken into account when measuring by AFM particularly soft interfaces such as living cells plasma membrane~\cite{Costa2014} or when unfolding single molecules~\cite{Rico2013}.

Moreover, the optomechanical interaction will not be constant if the alignment between the excitation laser and the AFM tip is not kept fixed during data acquisition, for instance, if the confocal spot is scanned and the tip is fixed as in the case of AFM coupled with a confocal microscope with a spinning disk operational scheme~\cite{Efremov2019} or with some STED microscopes.
While MLCT-BIO-DC minimizes optomechanical forces, quartz-like cantilevers such as qp-BioAC perform better for time-resolved fluorescence measurements due to lower tip luminescence.
We have shown that using a low excitation power of $120$~nW (more generally inferior to $10~\upmu$W) it is possible to image synchronously model membrane morphology, fluorescence intensity and lifetime minimizing optomechanical spurious forces ($<10$~pN) and tip and gold-coating on cantilever backside luminescence, therefore minimizing any tip-related artifact.

Concerning the nanopositioning experiment presented in Fig.~\ref{fig:force-lifetime}, the lifetime decrease occurs in absence of mechanical contact between the fluorescent probe and the gold substrate, at short distances where the AFM is not sensitive enough to detect any interaction force.
In perspective, this enables the possibility to obtain topographical images in liquid acquired in absence of AFM tip-sample mechanical contact.
It represents an important advance when it is to image soft, fluorescently-labeled, samples that are largely deformed by conventional AFM due to their extreme softness as it occurs in the case of living cell membranes. 

In our case, only a small drop ($7\%$ and $5\%$) in nanodiamond lifetime is observed in Figs.~\ref{fig:force-lifetime}c and d, respectively.
Although the change is small, uncertainty in lifetime characterization is in order of hundreds of picoseconds, therefore, the change observed is still significant.
Tisler~\emph{et al.} have observed a $35\%$ MIET efficiency when approaching a single NV~center $25$~nm nanodiamond attached to an AFM tip on top of graphene, a material that behaves like a metal~\cite{Tisler2013}.
In our case, the small MIET efficiency observed in Fig.~\ref{fig:force-lifetime} could be attributed to either the larger nanodiamond size ($\approx40$~nm) or the more abundant number of NV~centers ($\approx15$) that are randomly oriented in the nanodiamond core.
In the case of the experiment shown by Tisler~\emph{et al.}~\cite{Tisler2013} a complete lifetime spectroscopical change in respect to the nanodiamond-sample distance is not reported.
Indeed, in their case, the lifetime variation was measured only when the nanodiamond was in mechanical contact with graphene or the underlining substrate.

The lifetime/force curve, where the fluorescence decay is recorded for each $z$-position of the tip/emitter, enable us to gain insight in the near-field non-contact interactions, as it has been shown by Buchler~\emph{et al.} in the case of a gold nanoparticle approached to a glass coverslip~\cite{Buchler2005}, validating the theory predicted by Drexhage~\cite{Drexhage1970} in 1970.
Only when both lifetime and force signals are acquired, the precise mechanical contact point can be identified from the AFM interaction force and, consequently, an accurate fluorescent tip-sample distance can be measured. Indeed, the full control of the emitters nano-positioning is the key for future non-invasive (sub-piconewton forces) imaging modes, where the topography is mapped by relying on the quantum exchange of energy between donors (tip) and acceptors (sample) rather than the mechanical contact usual in AFM in liquid.

In summary, our results impose important experimental constraints for synchronous operation and interference-free correlative AFM and confocal time-resolved fluorescence.
In this work, we have quantified in pN/$\upmu$W the radiation pressure and photothermal detection forces acting on AFM cantilevers composed of different materials and in the case of different tip geometries.
While quartz-like AFM probes, minimizing tip luminescence, are more suited when fluorescence lifetime has to be measured, the overall optomechanical forces are minimized if radiation pressure and photothermal effect contributions are comparable, canceling each other.
However, our findings allow for the synchronous collection of AFM and FLIM images as well as for the simultaneous correlation of Force Spectroscopy (AFM-FS) with fluorescence lifetime.
We have shown the correlation of the topography (AFM) with the local viscosity (FLIM) in the presence of gel and fluid phase separated domains in model biological membranes labeled with a BODIPY fluorophore.
Finally, we have shown that the simultaneous correlation of force and fluorescence lifetime reveals a unique potential that we have used to reversibly tune the spectroscopic properties of a fluorescent nanodiamond attached to an AFM tip once it is placed in close vicinity with a gold substrate.

\section*{Materials and Methods} \label{mat}

\textbf{Atomic Force and Confocal Microscopies} images were acquired using a Nanowizard 4 (JPK Instruments, Bruker) equipped with a Tip Assisted Optics (TAO) module and a Vortis-SPM control unit.
The AFM cantilever detection system employs an infrared low-coherence light source, with emission centered at $980$~nm and further long-pass filtered at $850$~nm with a Schott RG850 filter.
The AFM head was mounted on a Zeiss inverted optical microscope.
A custom-made confocal microscope was coupled to the AFM using a Rock-PP supercontinuum laser (Leukos) as an excitation source operating at $20$~MHz ($50$~ns pulses of width $<100$~ps).
We used an oil immersion objective with a $1.4$~numerical aperture (Plan-Apochromat $100$x, FWD=$0.17$~mm, Zeiss) and a pinhole of $100~\upmu$m diameter size (P100D, Thorlabs).
We used an avalanche photodetector (SPCM-AQR-15, PerkinElmer) connected to an SPC-150 (Becker \& Hickl) TCSPC card to collect fluorescence signals.
The light source of the AFM detection system was filtered out in the fluorescence emission path using an ET800sp short pass filter (Chroma).
The excitation laser power was measured after the objective at the sample level with a S170C microscope slide power sensor and a PM100 energy meter interface (both purchased from Thorlabs).
The emission and excitation filters utilized are shown in Table~\ref{tab:filters}.
\begin{table}[hp!]
    \caption{Emission/Excitation filters.}
    \centering
    \begin{tabular}{cccc}
        \hline
    	\multicolumn{1}{l}{} & Model & Wavelength [nm] & Brand \\
    	\hline
    	\parbox[c]{2mm}{\multirow{2}{*}{\rotatebox[origin=c]{90}{Exc.{ }}}} & ZET488/10x & 448/10 & Chroma \\
    	& ZET532/10x & 532/10 & Chroma \\
    	\hline
    	\parbox[c]{2mm}{\multirow{2}{*}{\rotatebox[origin=c]{90}{Emiss.{ } }}} & FF01-525/39 & 525/39 & Semrock \\
    	& AT690/50m & 690/50 & Chroma \\
    	& FF01-709/167-25 & 709/167 & Semrock \\
    	\hline
    \end{tabular}
    \label{tab:filters}
\end{table}

During simultaneous AFM/FLIM images the AFM tip and confocal spot positions were fixed by co-aligning them.
Full alignment is obtained in two steps:
1. The tip is at first roughly aligned using white field illumination and then the alignment is fine-tuned using the increase of tip luminescence.
It is worth to notice that when tip and confocal spot are aligned, some AFM tips, such as qp-BioAC (Nanosensors), show colocalized tip luminescence and optical forces maxima, whereas so-called ``protruding AFM tips'' do not show maxima colocalization.
2. The sample is imaged and any mismatch between topography and FLIM image is adjusted by correcting the tip position.

Keeping the confocal spot and AFM tip aligned during data acquisition has the advantage to maintain constant all spurious effects resulting from synchronous operation.
In addition to the tip/cantilever luminescence background and the optomechanical deflection, we mention the mirror-like effect that is due to the reflection of the incoming excitation light by the metallic coating at the cantilever backside: this results in a higher excitation intensity that is constant all over the image acquisition if confocal spot and tip are aligned, whereas it is not in case of wide-field epifluorescence or confocal spinning disk operational schemes.

\textbf{Force/Lifetime curves.} Acquisition of approach-retract curves with simultaneous collection of force, photon intensity (photon counter) and fluorescent nanodiamond lifetime was performed using $532/10$~nm as excitation filter and $709/167$~nm as emission filter with a laser excitation power of $90~\upmu$W.
The number of photons emitted by the nanodiamond (>$6$~ns) and collected by the objective is in the order of tens of counts per second (cps), whereas normal acquisition speeds in modern AFM can reach hundreds of kHz~\cite{Amo2016}.
Since a hundreds or thousands of photons are needed to properly evaluate the lifetime, our acquisition bandwidth must be reduced to few tens/hundreds of Hz.
Faster acquisition rates would result in the collection of too few photons (or even none)  and the lifetime evaluation would be impossible.
For this reason we have fixed our acquisition bandwidth to $20$~Hz and performed forced curves in a quasi-static mode at a speed of $5$~nm/s.

Lifetime was evaluated by amplitude weighting the decay curve for times greater than $6$~ns.
All measurements were performed at room temperature and under liquid conditions in Dulbecco’s Phosphate Buffer Solution (DPBS), purchased from GIBCO, without MgCl$_2$ and CaCl$_2$ (GIBCO ref. 14190-094).
The tip was held constant and the sample was approached to the tip using the axial sample piezo of the TAO module.
The tip-confocal alignment was performed similarly to the method described above with a key change: we focused the confocal spot when the tip is in mechanical contact with the sample surface, hence the maximum signal from the nanodiamond/gold happens at the contact point  (Fig.~\ref{fig:force-lifetime}).

\textbf{AFM tips.} qp-BioAC (tip height $\approx 7~\mu$m) and ATEC-CONT (tip height $\approx 17~\mu$m) were purchased from Nanosensors and MLCT-BIO-DC (tip height $\approx 3.5~\mu$m) was purchased from Bruker.
MLCT-Bio-DC-C probes have a resonance frequency of $\approx 1$~kHz, ATEC-CONT of $\approx 6$~kHz and qp-BioAC of $\approx 25$~kHz and $\approx 45$~kHz (CB2 and CB1, respectively), all resonances measured in liquid environment.
Custom tips, presenting a plateau at the tip apex, were purchased from Nanoworld: they were fabricated modifying a qp-BioAC cantilever (see section 4 of supplementary information).
An electron beam deposited carbon tip was grown on top of a thin layer of gold.
Subsequently, a flat circular plateau of $\approx 50$~nm diameter was obtained at the tip apex by means of Field Ion Beam (FIB) to aid fluorescent nanodiamond fixation and stability.

For optomechanical force characterization, a dozen tips were used to fully characterize and understand the forces employed.
All probes of the same brand behaved similarly, with the same range of forces and same force/intensity distribution over the cantilever.
For the approach-retract experiment with the nanodiamonds, we have collected data from tens of different probes whereas each probe had a different nanodiamond that was carefully grafted.
In Fig.~\ref{fig:force-lifetime} two of such different tips were used.

\textbf{Tip calibration.} In all AFM experiments, the inverse optical lever sensitivity and lever stiffness of qp-BioAC and ATEC-CONT cantilevers were calibrated using a combination of a Sader~\cite{Sader2016} and thermal~\cite{Proksch2004} methods (called ``contact-free'' method in the JPK AFM instruments).
The fundamental peak was used with a correction of $0.817$ at ambient temperature and in liquid environment.

MLCT-BIO-DC triangular cantilever optical lever sensitivity was evaluated with the acquisition of an approach-retract curve (deflection versus $z$~scanner displacement curve) on a glass rigid surface and cantilevers stiffness were quantified consequently with a thermal method.

\textbf{Reverse-tip imaging.} Simultaneous intensity (photon counter) and force images presented in Fig.~\ref{fig:z-scan} were obtained by fixing the laser focus to $4\upmu$m from a glass coverslip and held in place throughout the measurement.
We employed the \textit{hover mode} (JPK instruments) where in the trace (forward) line the tip is in contact with the sample, whereas in the retrace (backward) line the tip is still following the topography (recorded from the trace) at a fixed distance from the sample defined by the AFM user.
In this way, in the retrace image, the deflection can be monitored and can be converted to the force acting on the cantilever at a specific distance.
By varying the latter, it is possible to probe the optomechanical interactions acting at different planes of the laser path.

Images were acquired using a $532/10$~nm excitation filter, a measured power of $300~\upmu$W and an emission filter of $690/50$~nm using $64$~lines $\times 64$~pixels, a line rate of $6$~Hz and scan size of $10~\upmu$m $\times 10~\upmu$m for ATEC-CONT probes, $15~\upmu$m $\times$ $15~\upmu$m for the MLCT-BIO-DC, and $30~\upmu$m $\times 30~\upmu$m for the qp-BioAC.
The photon counter signal from the APD was directly connected throught the input of the AFM electronics, not utilizing the TCSPC card, therefore, resulting in force and intensities images directly in the AFM software without need for any additional signal post-processing.
The intensity images units are given in counts per second (cps), which correspond to the photon collection rate by the AFM eletronics.

\textbf{Model membranes.} 1,2\hyp{}dipalmitoyl\hyp{}sn\hyp{}glycero\hyp{}3\hyp{}phosphocholine (DPPC) and 1,2\hyp{}dioleoyl\hyp{}sn\hyp{}glycero\hyp{}3\hyp{}phosphocholine (DOPC) were purchased from Avanti Polar Lipids Inc.
2\hyp{}(4,4\hyp{}difluoro\hyp{}5,7\hyp{}dimethyl\hyp{}4\hyp{}bora\hyp{}3a,4a\hyp{}diaza\hyp{}s\hyp{}indacene\hyp{}3\hyp{}dodecanoyl)\hyp{}1\hyp{}hexadecanoyl\hyp{}sn\hyp{}glycero\hyp{}3\hyp{}phosphocholine~(BODIPY) was purchased from Thermofisher.
Chloroform and methanol were purchased from Sigma‐Aldrich.
Membrane experiments were performed in DPBS buffer solution filtered before use with an inorganic membrane filter ($0.22~\upmu$m pore size, Whatman International Ltd).
DPPC and DOPC were individually dissolved in chloroform:methanol (v:v $1$:$1$) and supplemented with 0.1$\%$ BODIPY.
The solvent was then evaporated to dryness under nitrogen flow to obtain a thin film spread in a glass tube.
The dried lipids films were hydrated with DPBS buffer solution, previously heated at $60^\circ$C, until a final total concentration of $0.2$~mM.
The tube was later subjected to cycles of vortex mixing and heating at $60^\circ$C.
The vesicle suspension was extruded with a polycarbonate membrane filter ($100$~nm pore size, Whatman, purchased from Avanti Lipids).
Circular glass coverslips ($2.8$~cm diameter, $165~\upmu$m thick, purchased from Marienfeld) were cleaned by a cycle of sonication in KOH for 15 min, followed by a second cycle of sonication in deionized water for $15$~min.
Then, the glass coverslips were exposed to plasma (Expanded Plasma Cleaner PDC‐002, Harrick Scientific Corporation) at high RF power level for $15$~min.
Supported lipid bilayers (SLBs) were obtained by vesicles fusion method~\cite{Milhiet2001,Redondo-Morata2012,gumi2018plane}.
$300~\upmu$L of vesicles suspensions were deposited onto cleaned glass coverslips, previously mechanically fixed in an Attofluor chamber (Thermofisher), and incubated for $30$~min at $70^\circ$C.
Afterwards, the samples were rinsed several times with buffer solution to avoid unfused vesicles, always keeping the substrates hydrated and imaged by correlative confocal-AFM after $24$~hours.

\textbf{FLIM of model membranes.} Images were obtained using a $488/10$~nm excitation filter and a $535/39$~nm emission filter with a laser excitation power of $120$~nW.
AFM images of the membrane were acquired in contact mode using a rectangular qp-BioAC cantilever with a stiffness of $0.3$~N/m at constant force of $1.5$~nN at a line imaging rate of $0.25$~Hz with $128$~lines $\times$ $128$~pixels and with a scan size of $15~\upmu$m $\times 15~\upmu$m.
Images were collected with a sample scan with the AFM tip and excitation beam kept aligned and steady.
Lifetime was fitted with SPCimage (Becker \& Hickl) using a single exponential and spatial binning of~$1$.
AFM and FLIM images were synchronized by tagging each photon read by the TCSPC card with a TLL signal (pixel/line/frame) from the AFM instrument.

\textbf{Nanodiamonds and gold substrate.} Circular glass coverslips (Marienfeld) were subjected to sonication cycles in acetone ($5$~minutes) and ethanol ($5$~minutes), and subsequently dried under nitrogen flow.
$5$~nm or $10$~nm of gold were evaporated on top of coverslips by thermal evaporation method at a rate of $1\angstrom/s$ (AS053 Oerlikon thermal evaporating source).
Fluorescent Nanodiamonds (FNDs) were purchased from FND Biotech (Taiwan).
A solution of $40$~nm diameter nanodiamonds, each containing in average $15$~NV centers, at a concentration of $0.1$~mg/mL was sonicated for $15$~min and subsequently diluted to $0.1~\upmu$g/mL.
$5~\upmu$L were deposited on both gold-coated and glass coverslips and left to dry.
Finally, nanodiamonds were imaged by correlative confocal and AFM in DPBS solution.

\textbf{Nanodiamond grafting} at the AFM tip apex was performed on both ATEC-CONT and qp-BioAC tips.
Fluorescence images were acquired without AFM tip to pre-localize nanodiamond candidates on a glass coverslip for the grafting process.
Subsequently, an AFM image with low force setpoint ($\ll1$~nN) was acquired in the same area in AFM dynamic mode.
The PLL-coated tip was then pressed on the selected nanodiamond for $10$--$20$~seconds with a $1$~nN setpoint~\cite{Rondin2012}.
The use of PLL permanently sets the nanodiamond to be at the tip apex if low forces are used to avoid dislodging events.
The modified qp-BioAC, with a circular plateau at the tip apex, ensures higher nanodiamond stability. The absence of the plateau in the ATEC-CONT makes the nanodiamond more prone to dislodgement if higher forces are used.

\section*{Acknowledgments} \label{Acknowledgments}
L.C. and T.F. acknowledge support from CNRS Momentum program (2017).
O.S. acknowledges funding from the European Union's Horizon 2020 research and innovation program under the Marie Sklodowska-Curie grant agreement No. 721874 (SPM 2.0).
This project has received funding from the Agence Nationale de la Recherche (ANR) under grant agreement ANR-17-CE09-0026-02.
The CBS is a member of the France-BioImaging (FBI) and the French Infrastructure for the Integrated Biology (FRISBI), 2 national infrastructures supported by the French National Research Agency (ANR-10-INBS-04-01 and ANR-10-INBS-05 respectively).
The authors acknowledge Francesco Pedaci for discussions and Martina Di Muzio and the MicroFabSpace and Microscopy Characterization Facility, Unit 7 of ICTS “NANBIOSIS” from CIBER at IBEC, for providing the gold-coated coverslips.

\section*{Author contributions}
E.M, P.-E. M., L.C. and  T.F.D.F conceived the research.
T.F.D.F and L.C. conducted the experiments.
T.F.D.F analyzed the data and conceived the Python code for data analysis.
T.F.D.F, O.S., L.C. and E.M. set up the instrument.
All authors reviewed the manuscript.

\section*{Competing interests}
The authors declare no competing interests.

\bibliography{bib}

\begin{thebibliography}{10}
\urlstyle{rm}
\expandafter\ifx\csname url\endcsname\relax
  \def\url#1{\texttt{#1}}\fi
\expandafter\ifx\csname urlprefix\endcsname\relax\def\urlprefix{URL }\fi
\expandafter\ifx\csname doiprefix\endcsname\relax\def\doiprefix{DOI: }\fi
\providecommand{\bibinfo}[2]{#2}
\providecommand{\eprint}[2][]{\url{#2}}

\bibitem{Cheng2018}
\bibinfo{author}{Cheng, Y.}
\newblock \bibinfo{journal}{\bibinfo{title}{{Single-particle cryo-EM—How did
  it get here and where will it go}}}.
\newblock {\emph{\JournalTitle{Science}}} \textbf{\bibinfo{volume}{361}},
  \bibinfo{pages}{876--880}, \doiprefix\url{10.1126/science.aat4346}
  (\bibinfo{year}{2018}).

\bibitem{Willig2006}
\bibinfo{author}{Willig, K.~I.}, \bibinfo{author}{Rizzoli, S.~O.},
  \bibinfo{author}{Westphal, V.}, \bibinfo{author}{Jahn, R.} \&
  \bibinfo{author}{Hell, S.~W.}
\newblock \bibinfo{journal}{\bibinfo{title}{{STED microscopy reveals that
  synaptotagmin remains clustered after synaptic vesicle exocytosis}}}.
\newblock {\emph{\JournalTitle{Nature}}} \textbf{\bibinfo{volume}{440}},
  \bibinfo{pages}{935--939}, \doiprefix\url{10.1038/nature04592}
  (\bibinfo{year}{2006}).

\bibitem{Rust2006}
\bibinfo{author}{Rust, M.~J.}, \bibinfo{author}{Bates, M.} \&
  \bibinfo{author}{Zhuang, X.}
\newblock \bibinfo{journal}{\bibinfo{title}{{Sub-diffraction-limit imaging by
  stochastic optical reconstruction microscopy (STORM)}}}.
\newblock {\emph{\JournalTitle{Nature Methods}}} \textbf{\bibinfo{volume}{3}},
  \bibinfo{pages}{793--796}, \doiprefix\url{10.1038/nmeth929}
  (\bibinfo{year}{2006}).

\bibitem{Manley2008}
\bibinfo{author}{Manley, S.} \emph{et~al.}
\newblock \bibinfo{journal}{\bibinfo{title}{{High-density mapping of
  single-molecule trajectories with photoactivated localization microscopy}}}.
\newblock {\emph{\JournalTitle{Nature Methods}}} \textbf{\bibinfo{volume}{5}},
  \bibinfo{pages}{155--157}, \doiprefix\url{10.1038/nmeth.1176}
  (\bibinfo{year}{2008}).

\bibitem{balzarotti2017nanometer}
\bibinfo{author}{Balzarotti, F.} \emph{et~al.}
\newblock \bibinfo{journal}{\bibinfo{title}{Nanometer resolution imaging and
  tracking of fluorescent molecules with minimal photon fluxes}}.
\newblock {\emph{\JournalTitle{Science}}} \textbf{\bibinfo{volume}{355}},
  \bibinfo{pages}{606--612} (\bibinfo{year}{2017}).

\bibitem{Dufrene2017}
\bibinfo{author}{Dufr{\^{e}}ne, Y.~F.} \emph{et~al.}
\newblock \bibinfo{journal}{\bibinfo{title}{{Imaging modes of atomic force
  microscopy for application in molecular and cell biology}}}.
\newblock {\emph{\JournalTitle{Nature Nanotechnology}}}
  \textbf{\bibinfo{volume}{12}}, \bibinfo{pages}{295--307},
  \doiprefix\url{10.1038/nnano.2017.45} (\bibinfo{year}{2017}).

\bibitem{Ando2012}
\bibinfo{author}{Ando, T.}
\newblock \bibinfo{journal}{\bibinfo{title}{{High-speed atomic force microscopy
  coming of age}}}.
\newblock {\emph{\JournalTitle{Nanotechnology}}} \textbf{\bibinfo{volume}{23}},
  \bibinfo{pages}{062001}, \doiprefix\url{10.1088/0957-4484/23/6/062001}
  (\bibinfo{year}{2012}).

\bibitem{Nasrallah2019}
\bibinfo{author}{Nasrallah, H.} \emph{et~al.}
\newblock \bibinfo{title}{Imaging artificial membranes using high-speed atomic
  force microscopy}.
\newblock In \emph{\bibinfo{booktitle}{Atomic Force Microscopy}},
  \bibinfo{pages}{45--59} (\bibinfo{publisher}{Springer},
  \bibinfo{year}{2019}).

\bibitem{Kopek2012}
\bibinfo{author}{Kopek, B.~G.}, \bibinfo{author}{Shtengel, G.},
  \bibinfo{author}{Xu, C.~S.}, \bibinfo{author}{Clayton, D.~A.} \&
  \bibinfo{author}{Hess, H.~F.}
\newblock \bibinfo{journal}{\bibinfo{title}{{Correlative 3D superresolution
  fluorescence and electron microscopy reveal the relationship of mitochondrial
  nucleoids to membranes}}}.
\newblock {\emph{\JournalTitle{Proceedings of the National Academy of
  Sciences}}} \textbf{\bibinfo{volume}{109}}, \bibinfo{pages}{6136--6141},
  \doiprefix\url{10.1073/pnas.1121558109} (\bibinfo{year}{2012}).

\bibitem{Harke2012}
\bibinfo{author}{Harke, B.}, \bibinfo{author}{Chacko, J.~V.},
  \bibinfo{author}{Haschke, H.}, \bibinfo{author}{Canale, C.} \&
  \bibinfo{author}{Diaspro, A.}
\newblock \bibinfo{journal}{\bibinfo{title}{{A novel nanoscopic tool by
  combining AFM with STED microscopy}}}.
\newblock {\emph{\JournalTitle{Optical Nanoscopy}}}
  \textbf{\bibinfo{volume}{1}}, \bibinfo{pages}{3},
  \doiprefix\url{10.1186/2192-2853-1-3} (\bibinfo{year}{2012}).

\bibitem{Odermatt2015}
\bibinfo{author}{Odermatt, P.~D.} \emph{et~al.}
\newblock \bibinfo{journal}{\bibinfo{title}{{High-Resolution Correlative
  Microscopy: Bridging the Gap between Single Molecule Localization Microscopy
  and Atomic Force Microscopy}}}.
\newblock {\emph{\JournalTitle{Nano Letters}}} \textbf{\bibinfo{volume}{15}},
  \bibinfo{pages}{4896--4904}, \doiprefix\url{10.1021/acs.nanolett.5b00572}
  (\bibinfo{year}{2015}).

\bibitem{Dahmane2019a}
\bibinfo{author}{Dahmane, S.} \emph{et~al.}
\newblock \bibinfo{journal}{\bibinfo{title}{{Nanoscale organization of
  tetraspanins during HIV-1 budding by correlative dSTORM/AFM}}}.
\newblock {\emph{\JournalTitle{Nanoscale}}} \textbf{\bibinfo{volume}{11}},
  \bibinfo{pages}{6036--6044}, \doiprefix\url{10.1039/C8NR07269H}
  (\bibinfo{year}{2019}).

\bibitem{Gomez-Varela2020}
\bibinfo{author}{G{\'{o}}mez-Varela, A.~I.} \emph{et~al.}
\newblock \bibinfo{journal}{\bibinfo{title}{{Simultaneous co-localized
  super-resolution fluorescence microscopy and atomic force microscopy:
  combined SIM and AFM platform for the life sciences}}}.
\newblock {\emph{\JournalTitle{Scientific Reports}}}
  \textbf{\bibinfo{volume}{10}}, \bibinfo{pages}{1122},
  \doiprefix\url{10.1038/s41598-020-57885-z} (\bibinfo{year}{2020}).

\bibitem{Hillner1995}
\bibinfo{author}{Hillner, P.}, \bibinfo{author}{Walters, D.},
  \bibinfo{author}{Lal, R.}, \bibinfo{author}{Hansma, H.} \&
  \bibinfo{author}{Hansma, R.}
\newblock \bibinfo{journal}{\bibinfo{title}{{Combined Atomic Force and Confocal
  Laser Scanning Microscope}}}.
\newblock {\emph{\JournalTitle{Microscopy and Microanalysis}}}
  \textbf{\bibinfo{volume}{1}}, \bibinfo{pages}{127--130},
  \doiprefix\url{10.1017/S1431927695111277} (\bibinfo{year}{1995}).

\bibitem{Gradinaru2004}
\bibinfo{author}{Gradinaru, C.~C.}, \bibinfo{author}{Martinsson, P.},
  \bibinfo{author}{Aartsma, T.~J.} \& \bibinfo{author}{Schmidt, T.}
\newblock \bibinfo{journal}{\bibinfo{title}{{Simultaneous atomic-force and
  two-photon fluorescence imaging of biological specimens in vivo}}}.
\newblock {\emph{\JournalTitle{Ultramicroscopy}}}
  \textbf{\bibinfo{volume}{99}}, \bibinfo{pages}{235--245},
  \doiprefix\url{10.1016/j.ultramic.2003.12.009} (\bibinfo{year}{2004}).

\bibitem{Kassies2005}
\bibinfo{author}{Kassies, R.} \emph{et~al.}
\newblock \bibinfo{journal}{\bibinfo{title}{{Combined AFM and confocal
  fluorescence microscope for applications in bio-nanotechnology}}}.
\newblock {\emph{\JournalTitle{Journal of Microscopy}}}
  \textbf{\bibinfo{volume}{217}}, \bibinfo{pages}{109--116},
  \doiprefix\url{10.1111/j.0022-2720.2005.01428.x} (\bibinfo{year}{2005}).

\bibitem{Bek2008}
\bibinfo{author}{Bek, A.} \emph{et~al.}
\newblock \bibinfo{journal}{\bibinfo{title}{{Fluorescence Enhancement in Hot
  Spots of AFM-Designed Gold Nanoparticle Sandwiches}}}.
\newblock {\emph{\JournalTitle{Nano Letters}}} \textbf{\bibinfo{volume}{8}},
  \bibinfo{pages}{485--490}, \doiprefix\url{10.1021/nl072602n}
  (\bibinfo{year}{2008}).

\bibitem{Tisler2013}
\bibinfo{author}{Tisler, J.} \emph{et~al.}
\newblock \bibinfo{journal}{\bibinfo{title}{{Single Defect Center Scanning
  Near-Field Optical Microscopy on Graphene}}}.
\newblock {\emph{\JournalTitle{Nano Letters}}} \textbf{\bibinfo{volume}{13}},
  \bibinfo{pages}{3152--3156}, \doiprefix\url{10.1021/nl401129m}
  (\bibinfo{year}{2013}).

\bibitem{Tetienne2014b}
\bibinfo{author}{Tetienne, J.-p.}
\newblock \emph{\bibinfo{title}{{Un microscope de champ magn{\'{e}}tique
  bas{\'{e}} sur le d{\'{e}}faut azote-lacune du diamant: r{\'{e}}alisation et
  application {\`{a}} l'{\'{e}}tude de couches ferromagn{\'{e}}tiques
  ultraminces}}}.
\newblock Ph.D. thesis, \bibinfo{school}{{\'{E}}cole normale sup{\'{e}}rieure
  de Cachan - ENS Cachan} (\bibinfo{year}{2014}).

\bibitem{he2016simultaneous}
\bibinfo{author}{He, Y.}, \bibinfo{author}{Govind~Rao, V.},
  \bibinfo{author}{Cao, J.} \& \bibinfo{author}{Lu, H.~P.}
\newblock \bibinfo{journal}{\bibinfo{title}{Simultaneous spectroscopic and
  topographic imaging of single-molecule interfacial electron-transfer
  reactivity and local nanoscale environment}}.
\newblock {\emph{\JournalTitle{The journal of physical chemistry letters}}}
  \textbf{\bibinfo{volume}{7}}, \bibinfo{pages}{2221--2227},
  \doiprefix\url{10.1021/acs.jpclett.6b00862} (\bibinfo{year}{2016}).

\bibitem{Meller2006}
\bibinfo{author}{Meller, K.} \& \bibinfo{author}{Theiss, C.}
\newblock \bibinfo{journal}{\bibinfo{title}{{Atomic force microscopy and
  confocal laser scanning microscopy on the cytoskeleton of permeabilised and
  embedded cells}}}.
\newblock {\emph{\JournalTitle{Ultramicroscopy}}}
  \textbf{\bibinfo{volume}{106}}, \bibinfo{pages}{320--325},
  \doiprefix\url{10.1016/j.ultramic.2005.10.003} (\bibinfo{year}{2006}).

\bibitem{Laskowski2017}
\bibinfo{author}{Laskowski, P.~R.} \emph{et~al.}
\newblock \bibinfo{journal}{\bibinfo{title}{{High-Resolution Imaging and
  Multiparametric Characterization of Native Membranes by Combining Confocal
  Microscopy and an Atomic Force Microscopy-Based Toolbox}}}.
\newblock {\emph{\JournalTitle{ACS Nano}}} \textbf{\bibinfo{volume}{11}},
  \bibinfo{pages}{8292--8301}, \doiprefix\url{10.1021/acsnano.7b03456}
  (\bibinfo{year}{2017}).

\bibitem{Beicker2018}
\bibinfo{author}{Beicker, K.}, \bibinfo{author}{O'Brien, E.~T.},
  \bibinfo{author}{Falvo, M.~R.} \& \bibinfo{author}{Superfine, R.}
\newblock \bibinfo{journal}{\bibinfo{title}{{Vertical Light Sheet Enhanced
  Side-View Imaging for AFM Cell Mechanics Studies}}}.
\newblock {\emph{\JournalTitle{Scientific Reports}}}
  \textbf{\bibinfo{volume}{8}}, \bibinfo{pages}{1504},
  \doiprefix\url{10.1038/s41598-018-19791-3} (\bibinfo{year}{2018}).

\bibitem{Efremov2019}
\bibinfo{author}{Efremov, Y.~M.} \emph{et~al.}
\newblock \bibinfo{journal}{\bibinfo{title}{{Anisotropy vs isotropy in living
  cell indentation with AFM}}}.
\newblock {\emph{\JournalTitle{Scientific Reports}}}
  \textbf{\bibinfo{volume}{9}}, \bibinfo{pages}{5757},
  \doiprefix\url{10.1038/s41598-019-42077-1} (\bibinfo{year}{2019}).

\bibitem{Miranda2015}
\bibinfo{author}{Miranda, A.}, \bibinfo{author}{Martins, M.} \&
  \bibinfo{author}{{De Beule}, P. A.~A.}
\newblock \bibinfo{journal}{\bibinfo{title}{{Simultaneous differential spinning
  disk fluorescence optical sectioning microscopy and nanomechanical mapping
  atomic force microscopy}}}.
\newblock {\emph{\JournalTitle{Review of Scientific Instruments}}}
  \textbf{\bibinfo{volume}{86}}, \bibinfo{pages}{093705},
  \doiprefix\url{10.1063/1.4931064} (\bibinfo{year}{2015}).

\bibitem{Cazaux2016}
\bibinfo{author}{Cazaux, S.} \emph{et~al.}
\newblock \bibinfo{journal}{\bibinfo{title}{{Synchronizing atomic force
  microscopy force mode and fluorescence microscopy in real time for immune
  cell stimulation and activation studies}}}.
\newblock {\emph{\JournalTitle{Ultramicroscopy}}}
  \textbf{\bibinfo{volume}{160}}, \bibinfo{pages}{168--181},
  \doiprefix\url{10.1016/j.ultramic.2015.10.014} (\bibinfo{year}{2016}).

\bibitem{Sarkar2004}
\bibinfo{author}{Sarkar, A.}, \bibinfo{author}{Robertson, R.~B.} \&
  \bibinfo{author}{Fernandez, J.~M.}
\newblock \bibinfo{journal}{\bibinfo{title}{{Simultaneous atomic force
  microscope and fluorescence measurements of protein unfolding using a
  calibrated evanescent wave}}}.
\newblock {\emph{\JournalTitle{Proceedings of the National Academy of
  Sciences}}} \textbf{\bibinfo{volume}{101}}, \bibinfo{pages}{12882--12886},
  \doiprefix\url{10.1073/pnas.0403534101} (\bibinfo{year}{2004}).

\bibitem{Gumpp2009}
\bibinfo{author}{Gumpp, H.}, \bibinfo{author}{Stahl, S.~W.},
  \bibinfo{author}{Strackharn, M.}, \bibinfo{author}{Puchner, E.~M.} \&
  \bibinfo{author}{Gaub, H.~E.}
\newblock \bibinfo{journal}{\bibinfo{title}{{Ultrastable combined atomic force
  and total internal fluorescence microscope}}}.
\newblock {\emph{\JournalTitle{Review of Scientific Instruments}}}
  \textbf{\bibinfo{volume}{80}}, \bibinfo{pages}{063704},
  \doiprefix\url{10.1063/1.3148224} (\bibinfo{year}{2009}).

\bibitem{Ortega-Esteban2015}
\bibinfo{author}{Ortega-Esteban, A.} \emph{et~al.}
\newblock \bibinfo{journal}{\bibinfo{title}{{Fluorescence Tracking of Genome
  Release during Mechanical Unpacking of Single Viruses}}}.
\newblock {\emph{\JournalTitle{ACS Nano}}} \textbf{\bibinfo{volume}{9}},
  \bibinfo{pages}{10571--10579}, \doiprefix\url{10.1021/acsnano.5b03020}
  (\bibinfo{year}{2015}).

\bibitem{Uchihashi2016}
\bibinfo{author}{Uchihashi, T.}, \bibinfo{author}{Watanabe, H.},
  \bibinfo{author}{Fukuda, S.}, \bibinfo{author}{Shibata, M.} \&
  \bibinfo{author}{Ando, T.}
\newblock \bibinfo{journal}{\bibinfo{title}{{Functional extension of high-speed
  AFM for wider biological applications}}}.
\newblock {\emph{\JournalTitle{Ultramicroscopy}}}
  \textbf{\bibinfo{volume}{160}}, \bibinfo{pages}{182--196},
  \doiprefix\url{10.1016/j.ultramic.2015.10.017} (\bibinfo{year}{2016}).

\bibitem{Umakoshi2019}
\bibinfo{author}{Umakoshi, T.}, \bibinfo{author}{Fukuda, S.},
  \bibinfo{author}{Iino, R.}, \bibinfo{author}{Uchihashi, T.} \&
  \bibinfo{author}{Ando, T.}
\newblock \bibinfo{journal}{\bibinfo{title}{{High-speed near-field fluorescence
  microscopy combined with high-speed atomic force microscopy for biological
  studies}}}.
\newblock {\emph{\JournalTitle{Biochimica et Biophysica Acta (BBA) - General
  Subjects}}} \textbf{\bibinfo{volume}{1864}}, \bibinfo{pages}{0--1},
  \doiprefix\url{10.1016/j.bbagen.2019.03.011} (\bibinfo{year}{2020}).

\bibitem{Ramos2006}
\bibinfo{author}{Ramos, D.}, \bibinfo{author}{Tamayo, J.},
  \bibinfo{author}{Mertens, J.} \& \bibinfo{author}{Calleja, M.}
\newblock \bibinfo{journal}{\bibinfo{title}{{Photothermal excitation of
  microcantilevers in liquids}}}.
\newblock {\emph{\JournalTitle{Journal of Applied Physics}}}
  \textbf{\bibinfo{volume}{99}}, \bibinfo{pages}{124904},
  \doiprefix\url{10.1063/1.2205409} (\bibinfo{year}{2006}).

\bibitem{Favero2009}
\bibinfo{author}{Favero, I.} \& \bibinfo{author}{Karrai, K.}
\newblock \bibinfo{journal}{\bibinfo{title}{{Optomechanics of deformable
  optical cavities}}}.
\newblock {\emph{\JournalTitle{Nature Photonics}}}
  \textbf{\bibinfo{volume}{3}}, \bibinfo{pages}{201--205},
  \doiprefix\url{10.1038/nphoton.2009.42} (\bibinfo{year}{2009}).

\bibitem{Araghi2017}
\bibinfo{author}{Araghi, H.~Y.} \& \bibinfo{author}{Paige, M.~F.}
\newblock \bibinfo{journal}{\bibinfo{title}{{Insight into diacetylene
  photopolymerization in Langmuir-Blodgett films using simultaneous AFM and
  fluorescence microscopy imaging}}}.
\newblock {\emph{\JournalTitle{Surface and Interface Analysis}}}
  \textbf{\bibinfo{volume}{49}}, \bibinfo{pages}{1108--1114},
  \doiprefix\url{10.1002/sia.6284} (\bibinfo{year}{2017}).

\bibitem{YungHui2013}
\bibinfo{author}{{Yung Hui}, Y.} \emph{et~al.}
\newblock \bibinfo{journal}{\bibinfo{title}{{Tip-enhanced sub-diffraction
  fluorescence imaging of nitrogen-vacancy centers in nanodiamonds}}}.
\newblock {\emph{\JournalTitle{Applied Physics Letters}}}
  \textbf{\bibinfo{volume}{102}}, \bibinfo{pages}{013102},
  \doiprefix\url{10.1063/1.4773364} (\bibinfo{year}{2013}).

\bibitem{Rondin2012}
\bibinfo{author}{Rondin, L.} \emph{et~al.}
\newblock \bibinfo{journal}{\bibinfo{title}{{Nanoscale magnetic field mapping
  with a single spin scanning probe magnetometer}}}.
\newblock {\emph{\JournalTitle{Applied Physics Letters}}}
  \textbf{\bibinfo{volume}{100}}, \bibinfo{pages}{153118},
  \doiprefix\url{10.1063/1.3703128} (\bibinfo{year}{2012}).

\bibitem{silicon}
\bibinfo{author}{Li, Y.} \emph{et~al.}
\newblock \bibinfo{journal}{\bibinfo{title}{Broadband infrared
  photoluminescence in silicon nanowires with high density stacking faults}}.
\newblock {\emph{\JournalTitle{Nanoscale}}} \textbf{\bibinfo{volume}{7}},
  \bibinfo{pages}{1601--1605}, \doiprefix\url{10.1039/C4NR05410E}
  (\bibinfo{year}{2015}).

\bibitem{Wu2013}
\bibinfo{author}{Wu, Y.} \emph{et~al.}
\newblock \bibinfo{journal}{\bibinfo{title}{{Molecular rheometry: direct
  determination of viscosity in Lo and Ld lipid phases via fluorescence
  lifetime imaging}}}.
\newblock {\emph{\JournalTitle{Physical Chemistry Chemical Physics}}}
  \textbf{\bibinfo{volume}{15}}, \bibinfo{pages}{14986},
  \doiprefix\url{10.1039/c3cp51953h} (\bibinfo{year}{2013}).

\bibitem{SEANTIER2008326}
\bibinfo{author}{Seantier, B.}, \bibinfo{author}{Giocondi, M.-C.},
  \bibinfo{author}{Grimellec, C.~L.} \& \bibinfo{author}{Milhiet, P.-E.}
\newblock \bibinfo{journal}{\bibinfo{title}{Probing supported model and native
  membranes using afm}}.
\newblock {\emph{\JournalTitle{Current Opinion in Colloid \& Interface
  Science}}} \textbf{\bibinfo{volume}{13}}, \bibinfo{pages}{326 -- 337},
  \doiprefix\url{https://doi.org/10.1016/j.cocis.2008.01.003}
  (\bibinfo{year}{2008}).

\bibitem{Hof}
\bibinfo{author}{Przybylo, M.} \emph{et~al.}
\newblock \bibinfo{journal}{\bibinfo{title}{Lipid diffusion in giant
  unilamellar vesicles is more than 2 times faster than in supported
  phospholipid bilayers under identical conditions}}.
\newblock {\emph{\JournalTitle{Langmuir}}} \textbf{\bibinfo{volume}{22}},
  \bibinfo{pages}{9096--9099}, \doiprefix\url{10.1021/la061934p}
  (\bibinfo{year}{2006}).

\bibitem{Drexhage1970}
\bibinfo{author}{Drexhage, K.}
\newblock \bibinfo{journal}{\bibinfo{title}{{Influence of a dielectric
  interface on fluorescence decay time}}}.
\newblock {\emph{\JournalTitle{Journal of Luminescence}}}
  \textbf{\bibinfo{volume}{1-2}}, \bibinfo{pages}{693--701},
  \doiprefix\url{10.1016/0022-2313(70)90082-7} (\bibinfo{year}{1970}).

\bibitem{Costa2014}
\bibinfo{author}{Costa, L.} \emph{et~al.}
\newblock \bibinfo{journal}{\bibinfo{title}{{Spectroscopic Investigation of
  Local Mechanical Impedance of Living Cells}}}.
\newblock {\emph{\JournalTitle{PLoS ONE}}} \textbf{\bibinfo{volume}{9}},
  \bibinfo{pages}{e101687}, \doiprefix\url{10.1371/journal.pone.0101687}
  (\bibinfo{year}{2014}).

\bibitem{Rico2013}
\bibinfo{author}{Rico, F.}, \bibinfo{author}{Gonzalez, L.},
  \bibinfo{author}{Casuso, I.}, \bibinfo{author}{Puig-Vidal, M.} \&
  \bibinfo{author}{Scheuring, S.}
\newblock \bibinfo{journal}{\bibinfo{title}{{High-Speed Force Spectroscopy
  Unfolds Titin at the Velocity of Molecular Dynamics Simulations}}}.
\newblock {\emph{\JournalTitle{Science}}} \textbf{\bibinfo{volume}{342}},
  \bibinfo{pages}{741--743}, \doiprefix\url{10.1126/science.1239764}
  (\bibinfo{year}{2013}).

\bibitem{Buchler2005}
\bibinfo{author}{Buchler, B.~C.}, \bibinfo{author}{Kalkbrenner, T.},
  \bibinfo{author}{Hettich, C.} \& \bibinfo{author}{Sandoghdar, V.}
\newblock \bibinfo{journal}{\bibinfo{title}{{Measuring the Quantum Efficiency
  of the Optical Emission of Single Radiating Dipoles Using a Scanning
  Mirror}}}.
\newblock {\emph{\JournalTitle{Physical Review Letters}}}
  \textbf{\bibinfo{volume}{95}}, \bibinfo{pages}{063003},
  \doiprefix\url{10.1103/PhysRevLett.95.063003} (\bibinfo{year}{2005}).
\newblock \eprint{0506150}.

\bibitem{Amo2016}
\bibinfo{author}{Amo, C.~A.} \& \bibinfo{author}{Garcia, R.}
\newblock \bibinfo{journal}{\bibinfo{title}{{Fundamental High-Speed Limits in
  Single-Molecule, Single-Cell, and Nanoscale Force Spectroscopies}}}.
\newblock {\emph{\JournalTitle{ACS Nano}}} \textbf{\bibinfo{volume}{10}},
  \bibinfo{pages}{7117--7124}, \doiprefix\url{10.1021/acsnano.6b03262}
  (\bibinfo{year}{2016}).

\bibitem{Sader2016}
\bibinfo{author}{Sader, J.~E.} \emph{et~al.}
\newblock \bibinfo{journal}{\bibinfo{title}{{A virtual instrument to
  standardise the calibration of atomic force microscope cantilevers}}}.
\newblock {\emph{\JournalTitle{Review of Scientific Instruments}}}
  \textbf{\bibinfo{volume}{87}}, \bibinfo{pages}{093711},
  \doiprefix\url{10.1063/1.4962866} (\bibinfo{year}{2016}).

\bibitem{Proksch2004}
\bibinfo{author}{Proksch, R.}, \bibinfo{author}{Sch{\"{a}}ffer, T.~E.},
  \bibinfo{author}{Cleveland, J.~P.}, \bibinfo{author}{Callahan, R.~C.} \&
  \bibinfo{author}{Viani, M.~B.}
\newblock \bibinfo{journal}{\bibinfo{title}{{Finite optical spot size and
  position corrections in thermal spring constant calibration}}}.
\newblock {\emph{\JournalTitle{Nanotechnology}}} \textbf{\bibinfo{volume}{15}},
  \bibinfo{pages}{1344--1350}, \doiprefix\url{10.1088/0957-4484/15/9/039}
  (\bibinfo{year}{2004}).

\bibitem{Milhiet2001}
\bibinfo{author}{Milhiet, P.~E.}, \bibinfo{author}{Vi{\'{e}}, V.},
  \bibinfo{author}{Giocondi, M.-C.} \& \bibinfo{author}{{Le Grimellec}, C.}
\newblock \bibinfo{journal}{\bibinfo{title}{{AFM Characterization of Model
  Rafts in Supported Bilayers}}}.
\newblock {\emph{\JournalTitle{Single Molecules}}}
  \textbf{\bibinfo{volume}{2}}, \bibinfo{pages}{109--112},
  \doiprefix\url{10.1002/1438-5171(200107)2:2<109::AID-SIMO109>3.0.CO;2-L}
  (\bibinfo{year}{2001}).

\bibitem{Redondo-Morata2012}
\bibinfo{author}{Redondo-Morata, L.}, \bibinfo{author}{Giannotti, M.~I.} \&
  \bibinfo{author}{Sanz, F.}
\newblock \bibinfo{journal}{\bibinfo{title}{{Influence of Cholesterol on the
  Phase Transition of Lipid Bilayers: A Temperature-Controlled Force
  Spectroscopy Study}}}.
\newblock {\emph{\JournalTitle{Langmuir}}} \textbf{\bibinfo{volume}{28}},
  \bibinfo{pages}{12851--12860}, \doiprefix\url{10.1021/la302620t}
  (\bibinfo{year}{2012}).

\bibitem{gumi2018plane}
\bibinfo{author}{Gum{\'\i}-Audenis, B.} \emph{et~al.}
\newblock \bibinfo{journal}{\bibinfo{title}{In-plane molecular organization of
  hydrated single lipid bilayers: Dppc: cholesterol}}.
\newblock {\emph{\JournalTitle{Nanoscale}}} \textbf{\bibinfo{volume}{10}},
  \bibinfo{pages}{87--92}, \doiprefix\url{10.1039/C7NR07510C}
  (\bibinfo{year}{2018}).

\end{thebibliography}

\end{document}